\documentclass[apj]{emulateapj}
\usepackage{graphicx}
\shorttitle{AtomDB 2.0.2}
\shortauthors{Foster et al.}

\def\autos{\textsc{Autostructure}}
\def\ion#1#2{#1 \textsc{#2}}
\begin{document}

\title{Updated Atomic Data and Calculations for X-ray Spectroscopy}
\author{A. R. Foster, L. Ji\altaffilmark{1}, R. K. Smith and N. S. Brickhouse}
\affil{Smithsonian Astrophysical Observatory\\ Cambridge, MA 02144, USA}
\altaffiltext{1}{Purple Mountain Observatory, CAS, 2 West Beijing Rd, Nanjing, Jiangsu, P.R.China, 210008}
%\maketitle
\begin{abstract}
We describe the latest release of AtomDB, version 2.0.2, a database of atomic data and a plasma modeling code with a focus on X-ray astronomy. This release includes several major updates to the fundamental atomic structure and process data held within AtomDB, incorporating new ionization balance data, state-selective recombination data, and updated collisional excitation data for many ions, including the iron L-shell ions from Fe$^{+16}$ to Fe$^{+23}$ and all of the hydrogen- and helium-like sequences. We also describe some of the effects that these changes have on calculated emission and diagnostic line ratios, such as changes in the temperature implied by the He-like G-ratios of up to a factor of 2.
\end{abstract}

\section{Introduction}
X-ray spectra from astrophysical sources reveal both the constituent elements of those objects and the physics occurring within them. Successfully extracting this information from spectra requires both a model of the plasma and a large collection of data detailing the various atomic processes occurring within that plasma.

The recognition of the importance of the dielectronic recombination (DR) process for plasmas, even at high temperatures \citep{Burgess1964}, led to huge strides in modeling collisionally ionized, optically-thin astrophysical plasmas. The \cite{Cox1969} collection was one of the earliest attempts to collect atomic data for these models. Since then there have been a steady stream of refinements to both the collisional plasma models and the atomic data which underpin them. Different evolutions of such models have included those of \cite{Cox1971}, \cite{Mewe1972}, \cite{Landini1972}, \cite{Raymond1977}, and \cite{Brickhouse1995}.

Both the quantity and quality of relevant atomic data continue to grow: increases in computational power available allow for improved calculations, while advances in experimental methods and equipment allow more accurate measurements of a wider range of quantities. In modern analysis of collisionally ionized X-ray astrophysical spectra there are three widely-used atomic databases: SPEX v2.0 \citep{1996uxsa.conf..411K}, CHIANTI v7 \citep{Landi2012} and AtomDB v1.3.1 \citep{2001ApJ...556L..91S}. Each database has a slightly different focus: CHIANTI's main focus is on the EUV wavelengths for analyzing solar spectra, while SPEX and AtomDB focus on the X-ray ranges. Each of these databases continues to undergo periodic review and is updated as newer data become available; this paper introduces the release of AtomDB v2.0.2, describing the new data and improvements, how they have been implemented and what is planned for future releases.

\section{Calculation and Selection of Atomic Data}

\subsection{Priorities and Techniques}
In this paper we primarily discuss the theoretical sources of atomic data we have used for AtomDB v2.0.2. We will therefore begin by outlining a few of the guiding principles in our collection of data, and then briefly outlining the salient points of each major theoretical method. This is not intended as a comprehensive review of the atomic physics; for such a document see a text book such as \cite{Boyle2005}.

\subsubsection{Theoretical vs Experimental Data}
Accurate calculation of the emission from a collisionally-ionized plasma requires a broad range of atomic data, both in terms of the number of ions and processes included. The fundamental atomic data required can be split into structural and collisional categories. Structure data include the atomic energy levels and the radiative transition rates among them. Collisional data describe interactions between the ion and another particle in the plasma -- most commonly, an electron, although proton impact can have an effect on low-lying levels. The simplest atomic processes that can arise from collisions are electron impact excitation (EIE), electron impact ionization (EII), and radiative recombination (RR). More complex multi-step processes also occur, such as excitation-autoionization (EA) and dielectronic recombination (DR).

The ideal method for obtaining an accurate atomic database would be to have a large and accurate collection of experimental data covering all of these processes. In some cases, particularly for transition wavelengths, this can be done: the NIST database \citep{Ralchenko2011} collects wavelengths of line emission from a wide range of ions, from which highly accurate energy levels are calculated along with uncertainties. However for many processes, experimental measurements with sufficient breadth of coverage are unavailable: measuring collision strengths for one ion can take many months, while investigating it over the wide range of energies of astrophysical interest can take longer, or require a different apparatus. In addition, systematic errors can affect experimental measurement and are sometimes very difficult to remove. Finally, sometimes identifying what has been measured is not simple: a line with a wavelength may be observed, but attributing that to the correct energy levels requires some theoretical work and can be difficult in very complex ions. Therefore it is customary to use theoretical calculations in databases, with the available experimental information employed as a quality check.

One exception to this is line wavelengths: these can be measured with a high degree of accuracy. When using a fitting routine to match observed and modeled spectra, a small shift in wavelength can badly affect the results, inflating the fit statistic and occasionally causing other ion abundances to be affected. For this reason, observed wavelengths are preferred to the calculated wavelengths where they are available and the upper and lower levels can be accurately identified.

\subsubsection{Plasma Timescales}
The various processes in plasma occur on different timescales. A full discussion of these timescales is presented in \cite{Summers2006}, but it is sufficient here to note that $\tau_g \sim \tau_m \gg \tau_o \gg \tau_a \gg \tau_{ee}$, where timescales for change are: $\tau_g$ - the ground state ion population (typically due to ionization or recombination processes); $\tau_m$  - metastable states (those without a fast radiative decay to a lower level); $\tau_o$ - ordinary excited states (those with energy lower than the ionization potential of the ion); $\tau_a$ - auto-ionizing states (those with energy above the ionization potential, which can ionize spontaneously); and $\tau_{ee}$ - the thermalization timescale for electron-electron collisions to create a Maxwellian energy distribution in the plasma. A common assumption for astrophysical plasmas, and one that is made in the AtomDB database, is that the electrons have a Maxwellian energy distribution: Maxwellian-averaged collision strengths are stored for all ions. %Due to the differences in these timescales, the adjustments collisional radiative modeling \citep{bates} of plasmas concerns itself with the changes in plasma times

\subsubsection{Electron Impact Excitation}
Much of the new data compiled in v2.0.2 contains improvements in the EIE calculation. A variety of methods exist for calculating collisional excitation by electrons, in particular the distorted wave (DW) and $R$-matrix approaches. The DW technique is a perturbative approach, in which the electron is modeled as a wave perturbing the N or N+1 electron system.  In general, this approach works best for highly ionized targets (more than 2 or 3 times ionized), where the nucleus-electron potential can be treated as the dominant term, with all other interactions as lesser perturbations. It includes initial and final states of the system, but omits the many intermediate resonance states. These can be subsequently added back in in an isolated resonance approximation (i.e. assuming there is no correlation between each resonance). 

Close-coupled techniques, of which $R$-matrix \citep{Burke1971} is the most widely used for collisional excitation modeling, handle resonances naturally. In this model, the calculation is split into three regions based on the incoming electron-nuclear distance. The most significant change is in the inner region: the N+1 electron system wave functions are constructed from N and N+1 electron channels, with coupling possible with all the possible final states for the scattered electron. This allows modeling of resonances not as a separate process, but as an integral part of the calculation. It also allows $R$-matrix to be used to consistently provide ionization and, to some extent, recombination data. The inclusion of resonances, and some damping terms required by the technique to avoid over counting, enables the $R$-matrix method to give a more accurate treatment of lower energy collisions. It is, however, also a much more computationally intensive technique. For example, $R$-matrix calculations of H- and He-like systems (those with 1 and 2 electrons, respectively) rarely include levels above the $n=6$ shell, while DW calculations can include much higher levels with ease. There are some complex ions which are simply not possible to model well with $R$-matrix, though these ions are generally heavier than those of astrophysical interest. In the end, a trade-off is required between the accuracy of $R$-matrix and the wide coverage of ions and transitions offered by DW.

The $R$-matrix package has been modified and extended to include a variety of different models since its original introduction. The energy levels of heavier elements must be split in to their J-resolved sub levels for accurate spectroscopy. The Intermediate Coordinate Frame Transformation (ICFT, \citealt{Griffin1998}) method preforms much of the calculation in LS coupling, then applies a transform to the results to obtain IC coupled (and therefore J-resolved) data. This is beneficial due to the much lower number of energy terms in an LS coupled calculation compared to a similar IC problem; the computation time scales as approximately $n^3$, where $n$ is the number of levels or terms involved.

In Breit-Pauli $R$-matrix, outlined in \cite{Hummer1993}, the Hamiltonian for the system is formed by simplifying the fully-relativistic Dirac equation by treating the relativistic corrections as perturbations. For heavier elements ($Z>36$), the fully relativistic Dirac equation should be solved: the DARC \citep{Ait-Tahar1996} variant of $R$-matrix handles this, though it is usually not necessary for transitions from astrophysically relevant ions in the X-ray bandpass. Data from both the ICFT and Breit-Pauli methods have been included in this release.

For ions with low ionization energies, in particular near-neutral ions, loss of excited electrons to the continuum (i.e. ionization) can be a significant process and difficult to separate from excitation processes with near-ionization threshold energies. The introduction of \textit{pseudo-states} to represent the continuum states significantly improves collision cross section calculations, usually by reducing excessively high values. The $R$-matrix with Pseudo States (RMPS) code \citep{Bartschat1996} incorporates this, along with its relativistic counterpart DRMPS.

\subsection{Selection of Atomic Data}
In selecting the data for inclusion in this version of AtomDB, we have focused most attention on those areas where there was a known problem. For example if line ratios were consistently not matching observations, and no other explanation was satisfactory, we attempted to obtain updated rates. Hence we have focused on the He-like and Fe L-shell ions, where discrepancies have been noticed, and in extending H- and He-like ions to higher $n$-shells. The goal of the AtomDB project is not to generate atomic data, but to collect it from existing calculations and distribute the data to the X-ray astronomy community in useful formats. When there is no readily available published atomic data for a particular process or transition we will calculate and include it, however we will replace this calculated data with published results later when possible.

Finding multiple calculations of the same rates is relatively rare, but it does require that a decision be made as to which to include. In many cases, there is a clear improvement in the technique used in one of the calculations which means the results are likely to be significantly better than other data sets, such as comparing an $R$-matrix collision calculation with a DW one with obvious effects on the lower energy collision strengths. However, it is not as simple as using the latest data - using a newer technique for a calculation does not matter if it is not done well. We therefore look for comparisons with observation or laboratory measurements to select the best technique. Often multiple calculations have similar results, which makes the choice less relevant. There are relatively few cases where the data from two sources diverge strongly and there is little experimental data to provide any guidance; when these circumstances arise, we are forced to make a choice. We then follow the literature post-release to discover if there are any recurring issues.

\section{New Data}\label{sec:newdata}

Changes have been made throughout AtomDB. Several ions which were not previously included are now included in the database, while data for nearly all other ions have been improved by including more accurate calculations, by increasing the number of energy levels and transitions included for the ion, or a combination of both. In addition, major changes have been made to the ionization and, particularly, recombination data, with some significant changes to the resulting ionization balance and therefore to line emissivities. Here we outline each change in detail; in section \ref{sec:results} we will summarize the effect of these changes on the spectra observed by users. We note that in this release there are no changes to the format of the data files in which the database is stored; these are described in \cite{Smith2001}.

\subsection{Ionization and Recombination Rates}\label{sec:newdata:ionrec}

As experimental and theoretical methods improve, there are occasional efforts to collect the rates for ionization and recombination processes, including EII, EA, RR and DR recombination, for easy use by astrophysicists. Previous examples include the widely-used rates of \cite{Arnaud1992} and \cite{Mazzotta1998}: the latter was included in AtomDB v1.3.1. \cite{Bryans2006, Bryans2009} have produced a new compilation for elements from hydrogen ($Z=1$) to zinc ($Z$=30), which we have included in AtomDB v2.0.2. This provides a complete set of new rates for ionization and recombination for all astrophysically relevant ions from fully stripped up to Na-like systems (where Na-like refers to the recombining ion). Both the rates and an equilibrium ionization balance table are included in AtomDB v2.0.2.

In their papers \cite{Bryans2006, Bryans2009} compare the electron-impact ionization (EII) and recombination data with earlier data sets and with experimental measurements. The most significant changes, particularly for many L-shell ions, have been in the DR rates from the widely used \cite{Mazzotta1998} data. Calculations of DR rate coefficients by \cite{Badnell2006} and \cite{Gu2003, Gu2004} agree to within 35\% of each other. \cite{Bryans2006} also find that the DR rates agree to within 35\% for strong resonances when compared with results from storage ring experiments (see their paper for references). The exception to this is $\Delta n = 0$ core transitions at low temperature $(T \le 35,000K)$, where results diverge: in this temperature range single and doubly ionized ions are dominant and their populations should be treated with care, while at higher temperature no such complications arise. Due to the general good agreement of the methods, the DR data from \cite{Badnell2006} are used in the \cite{Bryans2006} compilation where possible as they cover a wider range of ions (the H-like through Na-like systems from He through to Zn). The remainder is from the earlier \cite{Mazzotta1998} data. The data set for DR into nitrogen-like ions \citep{Mitnik2004} have been revised by their authors since their publication and we have included this revised data here. The data sources for each isoelectronic sequence are listed in Table~\ref{tab:drref}.

RR rates have been taken from calculations by \cite{Badnell2006}. These have been compared with values from \cite{Gu2003a} and \cite{Verner1996}, and agree to within 10\% with the former, and 5\% with the latter. There is a larger divergence with the \cite{Gu2003a} data, up to 20\%, for H-like ions at temperatures in the upper range of their collisional ionization window, but in these cases the total recombination rate is dominated by the DR process so this effect is not important. Again due to the broader range of ions covered (bare nucleus to Na-like ions from H to Zn) the Badnell data set is adopted, with \cite{Mazzotta1998} data for the remaining ions.

For EII there are wider disagreements among data collections. The \cite{Dere2007} study collected ionization data from all ions of all elements from H to Zn, from both laboratory and theoretical sources. Another collection by \cite{Mattioli2007} covered a large subset of these ions: significant disagreement was found for many ions, with up to a factor of four difference. \cite{Bryans2009} use the \cite{Dere2007} data due to its broader scope, but the authors note (and we agree) that this discrepancy should be revisited and if possible resolved in the future.

\begin{deluxetable}{ll}
\tabletypesize{\small}
\tablecaption{Data sources for the DR state-selective rates.\label{tab:drref}}

\tablehead{\colhead{Sequence} & \colhead{Reference}}

\startdata
H-like & \cite{Badnell2006a} \\
He-like & \cite{Bautista2007}  \\
Li-like & \cite{Colgan2004} \\
Be-like & \cite{Colgan2003} \\
B-like & \cite{Altun2004} \\
C-like & \cite{Zatsarinny2004} \\
N-like & \cite{Mitnik2004} \\
O-like & \cite{Zatsarinny2003} \\
F-like & \cite{Zatsarinny2006} \\
Ne-like & \cite{Zatsarinny2004a} \\
Na-like & \cite{Altun2006} \\
Mg-like & \cite{Altun2007}
\enddata
\end{deluxetable}

\subsection{Final State Resolved DR \& RR Rates}\label{sec:newdata:ssdrrr}

The DR and RR data in the \cite{Bryans2006} collection contain total rate coefficients for recombination from the ground state of one ion to the next. The data on which these are based (Table~\ref{tab:drref}) are resolved into recombination into each excited state of the recombined ion. Inclusion of these rates when calculating the excited-state populations of the recombined ion will affect emission lines and diagnostic line ratios; they will do so even more in plasmas dominated by recombination. For this reason we have included state-selective data for recombination to all H-like, He-like and Li-like ions, and all iron ions from Fe$^{+16}$ to Fe$^{+22}$. As provided by \cite{Badnell2006}, the DR \& RR source data are split into level-resolved rates for capture into lower $n$ (typically $n \le 8$), with capture rates into higher $n$ provided as totals into each $n$-shell. Since the levels identified in the DR \& RR calculations and those in AtomDB are rarely identical, some level matching and cascade calculations have been required.

Using the  \autos\  \citep{Badnell1986} code, transition probabilities have been calculated for all the energy levels which appear in either the AtomDB data sets or the DR/RR calculations. In addition, transition probabilities between $n$ shells have been estimated using the hydrogenic approximation of \cite{Burgess1976}. Using this data, projection matrices have been formed to project the capture into the high-$n$ levels onto the lower-$n$ levels which are included in AtomDB. During the cascade, the electron density is assumed to be negligible and thus there is no collisional re-ionization/re-distribution. This cascaded population is then distributed according to the statistical weight of each level within the $n$ shell. The direct rate into each of the lower levels is then added to cascaded contribution, to obtain a total effective recombination rate to each level in the AtomDB database. %The end result of this process has been checked against the total rate provided in the original data set and found to be consistent, and is found to agree to within 2\%, which is acceptable considering the original data is provided to 2 decimal places.

This treatment of the cascade process is not ideal: firstly, the photons emitted during the cascade are not tracked and therefore are not included in any spectral models or cooling function estimates. For the purposes of AtomDB, this is relatively unimportant since most cascade emission will not be in the X-ray wavelengths, and the total omitted radiated power is small. In addition, the treatment of DR satellite lines is not consistent with these data: in AtomDB satellite lines are tabulated as a function of temperature obtained from separate calculations. The APEC code tracks the total DR rate from each approach to look for large discrepancies, with the average being a factor of 2 for strong  transitions, but there are outliers which are much larger. They would be better handled in a self-consistent manner during the original calculation of the DR rates.

\subsection{H-like Ions}\label{sec:newdata:hlike}

The data for hydrogenic ions have been upgraded in two major ways: (1) they have been extended to include higher $n$-shells (from $n=5$ to $n=10$) and (2) where available, data of a higher quality have been used for ions of astrophysical interest.

The existing data in AtomDB v1.3.1 for H-like ions come from a combination of the fully relativistic Dirac $R$-matrix calculations of \cite{Kisielius1996} for excitation from the ground, first excited and second excited state, and the non-relativistic calculations of \cite{Sampson1983} for the remainder. The \cite{Kisielius1996} data were only calculated for He and Fe; values for intermediate ions were obtained by scaling with $Z$. The maxmimum $n$-shell included in these data sets was $n$=5.

We have incorporated a new set of $R$-matrix results, obtained using the RMPS approach. When sophisticated techniques are applied to the H-like systems the results are not always published, as they are relatively straightforward calculations used as tests of the technique itself. As a result, $R$-matrix results could only be obtained for a few ions of astrophysical relevance. The ions for which $R$-matrix data were available and their sources are listed in Table~\ref{tab:hlike}.

\begin{deluxetable}{llll}
\tabletypesize{\small}
%\rotate
%\tablewidth{dimen}
%\tablenum{text}
%\tablecolumns{4}
\tablecaption{The sources of data and maximum $n$ shells for $R$-matrix collisional excitation calculations of H-like ions.\label{tab:hlike}}

\tablehead{\colhead{Ion} & \colhead{Reference} & \colhead{$n_{max}$} & 
\colhead{Coupling}}

\startdata
H$^{0}$ & \cite{Anderson2002} & 5 & LS \\
He$^{+}$ & \cite{Ballance2003} & 5 & LS \\
Li$^{+2}$ & \cite{Ballance2003} & 5 &  LS \\
Be$^{+3}$ & \cite{Ballance2003} & 5 &  LS \\
B$^{+4}$ & \cite{Ballance2003} & 5 &  LS \\
C$^{+5}$ & \cite{Ballance2003} & 5 &  LS \\
O$^{+7}$ & \cite{Ballance2003} & 5 &  LS \\
Ne$^{+9}$ & \cite{Ballance2003} & 6 &  LS \\
Fe$^{+25}$ & \cite{Ballance2003} & 5 &  IC
\enddata
\end{deluxetable}

For most of these data sets, calculations have been conducted in LS coupling. The resolution of these calculations stops at the term  (e.g. $^2P$), and not the level which includes fine structure splitting by total orbital angular momentum $J$ (e.g. $^2P_{\frac{1}{2}}, ^2P_{\frac{3}{2}}$). Due to the small $J$-splitting of the energy levels, term resolution is often adequate for hydrogen-like ions. For example, the splitting of the $n=2$ energy levels of \ion{Ne}{x} $\Delta E \approx 0.5$eV at $E \approx 1$keV. This would therefore require a spectral resolution of $E/\Delta E > 2000$ to resolve, while the Chandra HETG has only $E/\Delta E = 1000$. The resolution required to observe the lines varies as approximately $E/\Delta E \propto z^{-2}$, where $z$ is the atomic number of the element. At higher $z$ this splitting becomes apparent - for Fe XXVI, $E/\Delta E \approx 328$, which is observable in HETG spectra. This, in addition to the fact that AtomDB is used not only in analyzing existing mission data, but also for planning new missions which may require higher resolution spectra, requires all the atomic data in AtomDB to be resolved into its fine structure levels if possible.

We have therefore adopted a statistical splitting method for the hydrogenic ions to apply the fine-structure splitting: collisional excitation from \textit{ns} (e.g. 1s, 2s, 3s...) levels is taken from the term resolved $R$-matrix results: the rates are then split by statistical weights between J-resolved levels. This method is only possible for the \textit{ns} levels, yet this is largely sufficient for our needs since the ground and first excited state are \textit{ns} levels, and these levels are the dominant sources for excitation to higher levels and therefore the resulting spectra. This method was also adopted by the CHIANTI database.

For each H-like ion, the data available in the literature was expanded upon using a combination of FAC and \autos. Energy levels and transition rates were calculated for and between each level from $n=1, l=0$ up to $n=10, l=9$, for a total of 99 levels, using \autos. Energy levels calculated in this way agreed with values from NIST to within 0.05\%. Since observed wavelengths are stored separately from theoretical wavelengths in AtomDB and are used for spectral calculations in preference to theoretical values where possible, these were not corrected further during these calculation. Transition probabilities were calculated for all transitions up to the electric octupole and magnetic quadrupole.

Distorted wave Maxwell-averaged electron-impact collision strengths between each of these levels were calculated using FAC with the energy levels calculated by \autos. These ``fill in'' data sets were used in three ways: 

\begin{enumerate}
\item{To provide the collisional excitation and radiative rates for transitions where $10 \ge n_{upper} > 5$ (for $\mbox{Ne}^{+9}, 10 \ge n_{upper} > 6$)}
\item{To provide the collisional excitation and radiative rates between $nl$ levels, where $l_{lower} > 0 $}
\item{For those ions not included in Table \ref{tab:hlike}, to provide the full range of excitation and radiative rates for all transitions.}
\end{enumerate}

The $nl$ level transitions are weak and are not major contributors to the overall spectra, however they are included for completeness. Also for completeness, for all ions not mentioned in Table~\ref{tab:hlike} between H and Kr, the data obtained from this method were included in the database.

\subsection{He-like ions}
\label{sec:helike}
For the He-like ions, data have been incorporated from Intermediate Coupling Frame Transformation (ICFT) $R$-matrix calculations provided by A. Whiteford (2005, private communication) for all elements from $\mbox{C}^{5+}$ to $\mbox{Kr}^{35+}$. In this work, structure data is again generated by \autos, with the ICFT method being used for the collision strengths. A complete description of these calculations has not been published in the literature, but the data is available at the website of the UK APAP collaboration\footnote{ \url{http://amdpp.phys.strath.ac.uk/UK\_APAP/DATA/adf04/helike} }, and the technique is outlined by the author in \cite{Whiteford2001}.

\begin{figure}
\centering
\includegraphics[width=\linewidth]{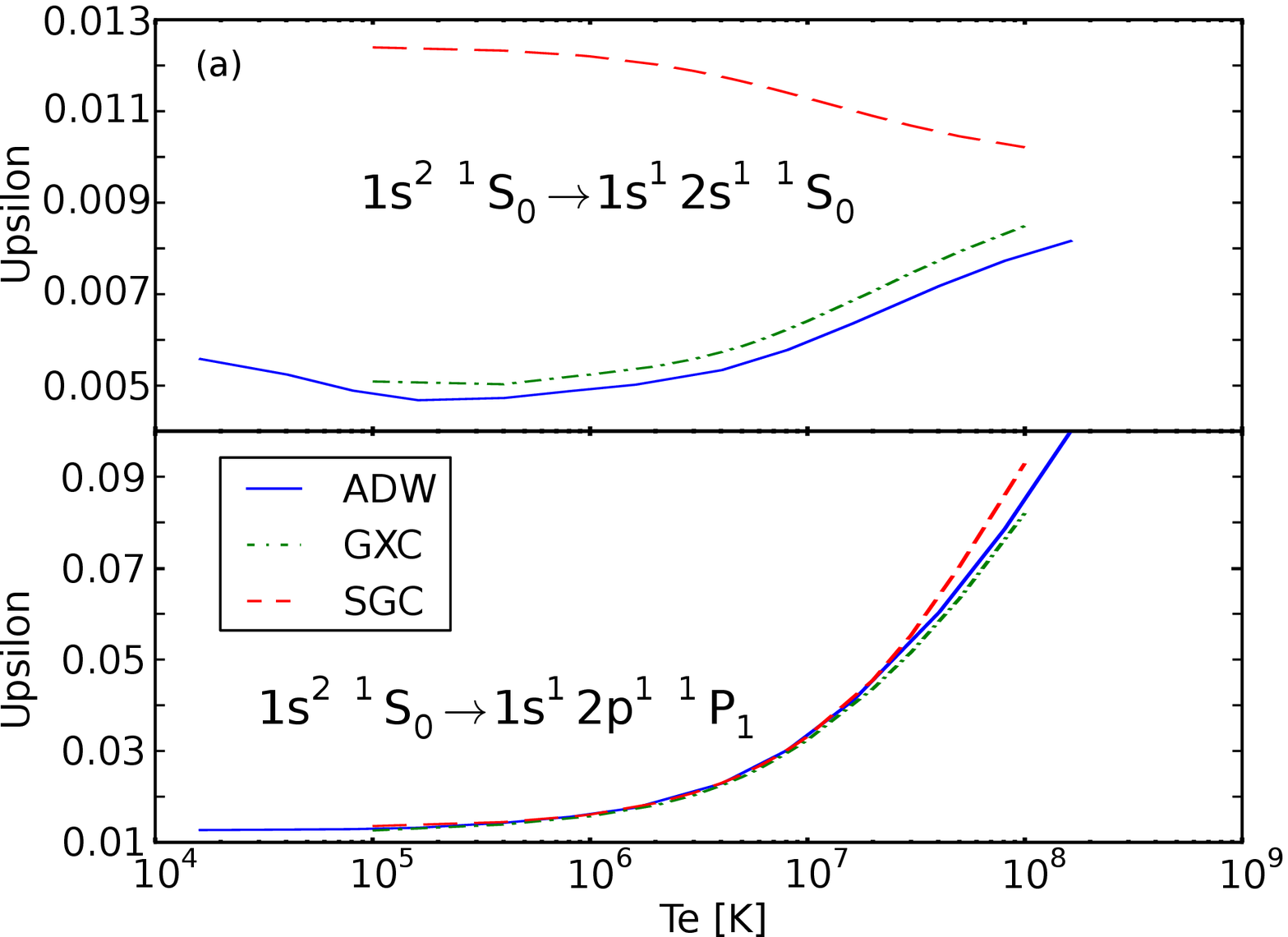}
\includegraphics[width=\linewidth]{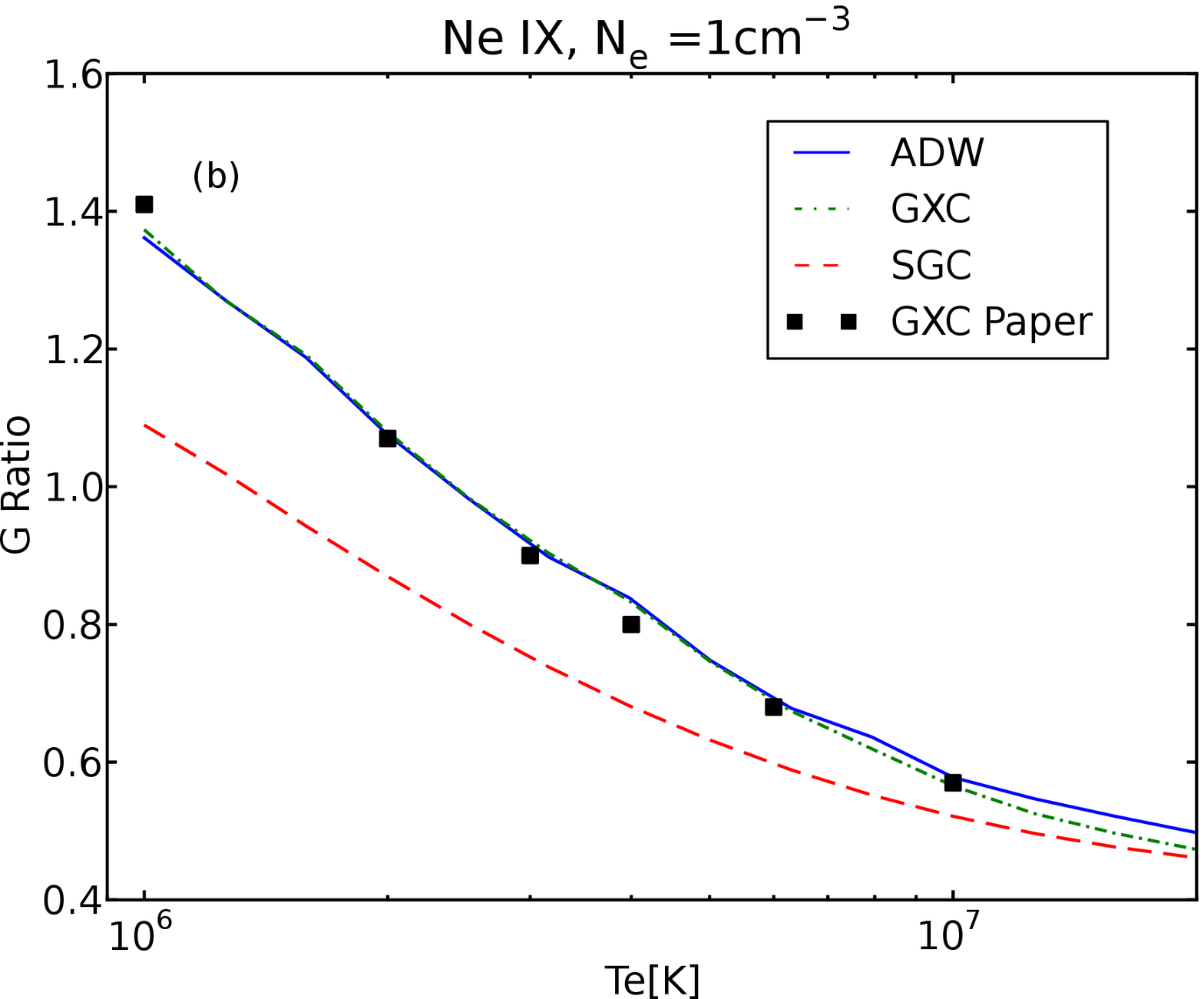}
\caption{\label{fig:hechenadw} The Maxwell-averaged collision strengths for two transitions of \ion{Ne}{ix}. \textit{Left} $1\mbox{s}^2 \, {}^1\mbox{S}_0 \rightarrow 1\mbox{s}^1 \, 2\mbox{s}^1 \, {}^1\mbox{S}_0$, and \textit{right} $1\mbox{s}^2 \, {}^1\mbox{S}_0 \rightarrow 1\mbox{s}^1 \, 2\mbox{p}^1 \, {}^1\mbox{P}_1$. The dotted line (SGC) is the data of \cite{Sampson1983}, used in AtomDB v1.3.1, the solid line (ADW) denotes the data of A. Whiteford (2003, private communication), the dash-dot line (GXC) is that of \cite{Chen2006}. The lower panel shows the G-ratio for both all three calculations, including recombination cascades. The squares are the values quoted for G-given in \cite{Chen2006}, our recalculation differs slightly due to different handling of the recombination and cascades.}
\end{figure}

Since this data is an unpublished collection, we have compared it with the fully relativistic $R$-matrix calculation of collisional excitation for \ion{Ne}{ix} by \citep{Chen2006}. The Maxwell-averaged collision strengths were found to agree with each other to within 5\% for all transitions, with the exception of those from the $1\mbox{s}^2~^1\mbox{S}_0 \to 1\mbox{s}^1~n\mbox{s}^1~^1\mbox{S}_0$, where the agreement is still within 10\%. A best case and worst case scenario comparison is shown in Figure~\ref{fig:hechenadw}. Also included in this figure are the collision strengths from \cite{Sampson1983}, which were used in AtomDB v1.3.1. These data were obtained from non-relativistic Coulomb Born Exchange calculations, and only at 9 energies such that resonance effects are not included.

There are two widely used diagnostic line ratios which arise from the $n=2\rightarrow 1$ transitions of He-like systems: the electron temperature sensitive G-ratio, and the electron density sensitive R-ratio \citep{Gabriel1969}. In the notation of  \cite{Gabriel1969}, these lines of interest are: the resonance line, $w$, $(1\mbox{s}\ 2\mbox{p}\ {}^1\mbox{P}_1\rightarrow \mbox{ground})$; the intercombination lines, $x$ and $y$, $(1\mbox{s}\ 2\mbox{p}\ {}^3\mbox{P}_{[1,2]}\rightarrow \mbox{ground})$; and the forbidden line, $z$, $(1\mbox{s}\ 2\mbox{s}\ {}^3\mbox{S}_{1}\rightarrow \mbox{ground})$. The G-ratio, $(x+y+z)/w$  decreases as the electron temperature, $T_e$, increases, allowing a this line ratios to be used to measure $T_e$. The decrease is driven by a number of factors: the excitation of the forbidden level has strong low energy resonances, and is therefore more pronounced at low temperature. In addition, the most prominent collisional excitation process changes: to excite the triplet levels of the $x, y$ and $z$ lines, the incident electron must exchange with an electron in the target ion, while the singlet level can be populated directly by excitation. The cross section for the former process is larger at lower energy and declines at higher energies while the reverse is true for direct excitation. The resonance line does not completely dominate due to recombination and direct excitation into higher energy levels, and the subsequent cascade to the $n=2$ shell, which populates them broadly in line with their statistical weights. This in turn prevents the G-ratio from falling significantly below 0.5.

The R-ratio, $z/(x+y)$, is sensitive to the electron density, $N_e$. At very low densities, electrons entering the metastable $1\mbox{s}\ 2\mbox{s}\ ^3\mbox{S}_{1}$ level decay to the ground state despite the state's very long lifetime, as there is no alternative lower energy state to decay to. At higher densities, electron collisions depopulate this state, leading to a relative decline of the forbidden line. This makes the R-ratio density sensitive, while also being relatively insensitive to temperature.

In the lower panel of Figure~\ref{fig:hechenadw} we compare the G-ratio from the different \ion{Ne}{ix} calculations for electron impact excitation. The two $R$-matrix methods are very similar in their results, showing that relativistic corrections are not large for \ion{Ne}{ix}. The similarity between the fully- and partially-relativistic $R$-matrix data is not unexpected for the Ne-like ions. When approaching heavier, more highly ionized ions (e.g. \ion{Fe}{xxv}), relativistic effects will become more significant and a larger discrepancy is expected. In \cite{Chen2008}, the authors compare their results with EBIT measurements from \cite{Wargelin1993} and find very good agreement. This differs significantly from the results of \cite{Sampson1983}. We have incorporated the data of A. Whiteford for the He-like isoelectronic sequence into AtomDB v2.0.2: it provides a high-quality uniform data set covering a wide range of ions. 

As with the H-like ions we have extended the range of these calculations using FAC to obtain collision strengths for collisions with $5 < n \le 10$. This was done using exactly the same method as for the H-like ions.

\subsection{Iron L-shell Data}
The data for the iron L-shell ions ($\mbox{Fe}^{+16}$ to $\mbox{Fe}^{+23}$) in AtomDB v1.3.1 is taken from a series of HULLAC \citep{2001JQSRT..71..169B} calculations by D. Liedahl (1997, private communication). While these data have never been explicitly published, the data for $\mbox{Fe}^{+19}$ to $\mbox{Fe}^{+23}$ are described in \cite{Liedahl1995} and \cite{Wargelin1998}. In summary, the structure was calculated using the relativistic, muticonfiguration parametric potential method \citep{Klapisch1971, Klapisch1977}, while the electron collision data are semi-relativistic distorted wave calculations, and include excitation and radiative decay from up to the $n=5$ to $n=7$ shell, depending on the ion.

Significant advances have been made in collisional excitation calculations since the release of AtomDB v1.3.1. These advances are largely due to the Iron Project, initiated by \cite{Hummer1993}. We have taken advantage of this data to upgrade the collision strength data for these ions to $R$-matrix calculations. Table~\ref{tab:fe_collision} lists the references for each ion in the Fe L-shell series. These new calculations include, in general, fewer configurations than the existing HULLAC data; as a result the newer data have been merged with the existing data, updating or adding as appropriate. Much of the older data remains for transitions not included in the newer data. Table \ref{tab:fe_configs}\ shows the configurations included for each ion in both the new and old datasets; unless otherwise noted, a level included in the new $R$-matrix data is also included in the old HULLAC data. In general the HULLAC data includes $n$ shells two or three higher than those in the $R$-matrix data.
\begin{deluxetable}{lll}
%\tabletypesize{\small}
%\rotate
%\tablewidth{dimen}
%\tablenum{text}
%\tablecolumns{4}
\tablecaption{The data sources used for iron L-shell ions, and the maximum $n$ shell included for each.\label{tab:fe_collision}}

\tablehead{\colhead{Ion} & \colhead{Reference} & \colhead{$n_{max}$}}

\startdata
Fe$^{+16}$ & \cite{Loch2006} & 4 \\
Fe$^{+17}$ & \cite{Witthoeft2007} & 4 \\
Fe$^{+18}$ & \cite{Butler2008} & 4 \\
Fe$^{+19}$ & \cite{Witthoeft2007a} & 4 \\
Fe$^{+20}$ & \cite{Badnell2001} & 4 \\
Fe$^{+21}$ & \cite{Badnell2001a} & 4 \\
Fe$^{+22}$ & \cite{Chidichimo2005}\tablenotemark{a}\tablenotetext{a}{as amended by \cite{DelZanna2005}} & 4 \\
Fe$^{+23}$ & \cite{Whiteford2002} & 3
\enddata

\end{deluxetable}

\begin{deluxetable*}{lll}
\tablecaption{The configurations included in the iron L-shell datasets listed in Table \ref{tab:fe_collision}, and the additional configurations included in the AtomDB v1.3.1 HULLAC datasets.\label{tab:fe_configs}}

\tablehead{\colhead{Ion} & \colhead{Configurations (New data)} & \colhead{Additional Configurations (HULLAC data)}}

\startdata
Fe$^{+16}$ & 
$\mbox{1s}^{2} \mbox{2s}^{2} \mbox{2p}^{6}$,
$\mbox{1s}^{2} \mbox{2s}^{2} \mbox{2p}^{5} \mbox{3l}^{1}$,
$\mbox{1s}^{2} \mbox{2s}^{1} \mbox{2p}^{6} \mbox{3l}^{1}$,&
$\mbox{1s}^{2} \mbox{2s}^{2} \mbox{2p}^{5} \mbox{6l}^{1}$,
$\mbox{1s}^{2} \mbox{2s}^{2} \mbox{2p}^{5} \mbox{7l}^{1}$,
$\mbox{1s}^{2} \mbox{2s}^{1} \mbox{2p}^{6} \mbox{5l}^{1}$,\\

& 
$\mbox{1s}^{2} \mbox{2s}^{2} \mbox{2p}^{5} \mbox{4l}^{1}$,
$\mbox{1s}^{2} \mbox{2s}^{2} \mbox{2p}^{5} \mbox{5l}^{1}$,
$\mbox{1s}^{2} \mbox{2s}^{1} \mbox{2p}^{6} \mbox{4l}^{1}$ &
$\mbox{1s}^{2} \mbox{2s}^{1} \mbox{2p}^{6} \mbox{6l}^{1}$,
$\mbox{1s}^{2} \mbox{2s}^{1} \mbox{2p}^{6} \mbox{7l}^{1}$ \\

Fe$^{+17}$ &
$\mbox{1s}^{2} \mbox{2s}^{2} \mbox{2p}^{5}$,
$\mbox{1s}^{2} \mbox{2s}^{1} \mbox{2p}^{6}$,
$\mbox{1s}^{2} \mbox{2s}^{2} \mbox{2p}^{4} \mbox{3l}^{1}$,&
$\mbox{1s}^{2} \mbox{2p}^{6} \mbox{3l}^{1}$,
$\mbox{1s}^{2} \mbox{2s}^{1} \mbox{2p}^{5} \mbox{4l}^{1}$,
$\mbox{1s}^{2} \mbox{2s}^{2} \mbox{2p}^{4} \mbox{5l}^{1}$,\\

&
$\mbox{1s}^{2} \mbox{2s}^{1} \mbox{2p}^{5} \mbox{3l}^{1}$,
$\mbox{1s}^{2} \mbox{2s}^{2} \mbox{2p}^{4} \mbox{4l}^{1}$ &
$\mbox{1s}^{2} \mbox{2s}^{1} \mbox{2p}^{5} \mbox{5l}^{1}$,
$\mbox{1s}^{2} \mbox{2p}^{6} \mbox{4l}^{1}$,
$\mbox{1s}^{2} \mbox{2p}^{6} \mbox{5l}^{1}$ \\

Fe$^{+18}$ & 
$\mbox{1s}^{2} \mbox{2s}^{2} \mbox{2p}^{4}$,
$\mbox{1s}^{2} \mbox{2s}^{1} \mbox{2p}^{5}$,
$\mbox{1s}^{2} \mbox{2p}^{6}$,&
$\mbox{1s}^{2} \mbox{2s}^{1} \mbox{2p}^{4} \mbox{4l}^{1}$,
$\mbox{1s}^{2} \mbox{2s}^{2} \mbox{2p}^{3} \mbox{5l}^{1}$,
$\mbox{1s}^{2} \mbox{2p}^{5} \mbox{4l}^{1}$,\\

&
$\mbox{1s}^{2} \mbox{2s}^{2} \mbox{2p}^{3} \mbox{3l}^{1}$,
$\mbox{1s}^{2} \mbox{2s}^{1} \mbox{2p}^{4} \mbox{3l}^{1}$,
$\mbox{1s}^{2} \mbox{2p}^{5} \mbox{3l}^{1}$,&
$\mbox{1s}^{2} \mbox{2s}^{1} \mbox{2p}^{4} \mbox{5l}^{1}$,
$\mbox{1s}^{2} \mbox{2p}^{5} \mbox{5l}^{1}$ \\

&
$\mbox{1s}^{2} \mbox{2s}^{2} \mbox{2p}^{3} \mbox{4l}^{1}$
&\\

Fe$^{+19}$ & 
$\mbox{1s}^{2} \mbox{2s}^{2} \mbox{2p}^{3}$,
$\mbox{1s}^{2} \mbox{2s}^{1} \mbox{2p}^{4}$,
$\mbox{1s}^{2} \mbox{2p}^{5}$, &
$\mbox{1s}^{2} \mbox{2p}^{4} \mbox{3l}^{1}$,
$\mbox{1s}^{2} \mbox{2s}^{1} \mbox{2p}^{3} \mbox{4l}^{1}$,
$\mbox{1s}^{2} \mbox{2s}^{2} \mbox{2p}^{2} \mbox{5l}^{1}$,\\
&
$\mbox{1s}^{2} \mbox{2s}^{2} \mbox{2p}^{2} \mbox{3l}^{1}$,
$\mbox{1s}^{2} \mbox{2s}^{1} \mbox{2p}^{3} \mbox{3l}^{1}$,
$\mbox{1s}^{2} \mbox{2s}^{2} \mbox{2p}^{2} \mbox{4l}^{1}$&
$\mbox{1s}^{2} \mbox{2p}^{4} \mbox{4l}^{1}$,
$\mbox{1s}^{2} \mbox{2s}^{1} \mbox{2p}^{3} \mbox{5l}^{1}$ \\

Fe$^{+20}$ & 
$\mbox{1s}^{2} \mbox{2s}^{2} \mbox{2p}^{2}$,
$\mbox{1s}^{2} \mbox{2s}^{1} \mbox{2p}^{3}$,
$\mbox{1s}^{2} \mbox{2p}^{4}$,&
$\mbox{1s}^{2} \mbox{2p}^{3} \mbox{3p}^{1}$,
$\mbox{1s}^{2} \mbox{2p}^{3} \mbox{3d}^{1}$,
$\mbox{1s}^{2} \mbox{2s}^{1} \mbox{2p}^{2} \mbox{4l}^{1}$,\\

&$\mbox{1s}^{2} \mbox{2s}^{2} \mbox{2p}^{1} \mbox{3l}^{1}$,
$\mbox{1s}^{2} \mbox{2s}^{1} \mbox{2p}^{2} \mbox{3l}^{1}$,
$\mbox{1s}^{2} \mbox{2p}^{3} \mbox{3s}^{1}$,&
$\mbox{1s}^{2} \mbox{2s}^{2} \mbox{2p}^{1} \mbox{5l}^{1}$,
$\mbox{1s}^{2} \mbox{2p}^{3} \mbox{4l}^{1}$,
$\mbox{1s}^{2} \mbox{2s}^{1} \mbox{2p}^{2} \mbox{5l}^{1}$ \\

&$\mbox{1s}^{2} \mbox{2s}^{2} \mbox{2p}^{1} \mbox{4l}^{1}$ &\\

Fe$^{+21}$ & 
$\mbox{1s}^{2} \mbox{2s}^{2} \mbox{2p}^{1}$,
$\mbox{1s}^{2} \mbox{2s}^{1} \mbox{2p}^{2}$,
$\mbox{1s}^{2} \mbox{2p}^{3}$, &
$\mbox{1s}^{2} \mbox{2s}^{2} \mbox{5l}^{1}$,
$\mbox{1s}^{2} \mbox{2s}^{1} \mbox{2p}^{1} \mbox{5l}^{1}$,
$\mbox{1s}^{2} \mbox{2s}^{2} \mbox{6l}^{1}$,\\

&$\mbox{1s}^{2} \mbox{2s}^{2} \mbox{3l}^{1}$,
$\mbox{1s}^{2} \mbox{2s}^{1} \mbox{2p}^{1} \mbox{3l}^{1}$,
$\mbox{1s}^{2} \mbox{2s}^{2} \mbox{4l}^{1}$,
&$\mbox{1s}^{2} \mbox{2s}^{1} \mbox{2p}^{1} \mbox{6l}^{1}$\\

&$\mbox{1s}^{2} \mbox{2s}^{1} \mbox{2p}^{1} \mbox{4l}^{1}$,
$\mbox{1s}^{2} \mbox{2p}^{2} \mbox{3l}^{1}\tablenotemark{a}$ &\\

Fe$^{+22}$ & 
$\mbox{1s}^{2} \mbox{2s}^{2}$,
$\mbox{1s}^{2} \mbox{2s}^{1} \mbox{2p}^{1}$,
$\mbox{1s}^{2} \mbox{2p}^{2}$,&
$\mbox{1s}^{2} \mbox{2s}^{1} \mbox{5l}^{1}$,
$\mbox{1s}^{2} \mbox{2p}^{1} \mbox{5l}^{1}$,
$\mbox{1s}^{2} \mbox{2s}^{1} \mbox{6l}^{1}$,\\

&$\mbox{1s}^{2} \mbox{2s}^{1} \mbox{3l}^{1}$,
$\mbox{1s}^{2} \mbox{2p}^{1} \mbox{3l}^{1}$,
$\mbox{1s}^{2} \mbox{2s}^{1} \mbox{4l}^{1}$,&
$\mbox{1s}^{2} \mbox{2p}^{1} \mbox{6l}^{1}$,
$\mbox{1s}^{2} \mbox{2s}^{1} \mbox{7l}^{1}$,
$\mbox{1s}^{2} \mbox{2p}^{1} \mbox{7l}^{1}$\\

&$\mbox{1s}^{2} \mbox{2p}^{1} \mbox{4l}^{1}$ &\\

Fe$^{+23}$ & 
$\mbox{1s}^{2} \mbox{2l}^{1}$,
$\mbox{1s}^{2} \mbox{3l}^{1}$,
$\mbox{1s}^{2} \mbox{4l}^{1}$,&
$\mbox{1s}^{2} \mbox{6l}^{1}$,
$\mbox{1s}^{2} \mbox{7l}^{1}$\\

&$\mbox{1s}^{2} \mbox{5l}^{1}$,
$\mbox{1s}^{1} \mbox{2s}^{2}\tablenotemark{a}$
$\mbox{1s}^{1} \mbox{2s}^{1} \mbox{2p}^{1}\tablenotemark{a}$,&\\

&$\mbox{1s}^{1} \mbox{2p}^{2}\tablenotemark{a}$,
$\mbox{1s}^{1} \mbox{2s}^{1} \mbox{3l}^{1}\tablenotemark{a}$,
$\mbox{1s}^{1} \mbox{2p}^{1} \mbox{3l}^{1}\tablenotemark{a}$ &
\enddata
\tablenotetext{a}{These configurations were included in the new datasets but were not in the HULLAC dataset from AtomDB v1.3.1}
\end{deluxetable*}

Merging the two data sets requires a method for matching levels between the two different data sets. This is complicated by the different coupling schemes reported for both data sets: the Liedahl HULLAC data is \textit{jj} coupled, while the new $R$-matrix data is \textit{LS} coupled. Levels have been matched by grouping levels with the same electron shell occupancies and $J$ quantum number, sorting these into energy order, and linking it to its counterpart in the other data set after the performing the same process. Where level energies disagree, the $R$-matrix calculations have been used. Spot checks on many strong lines show that these have been correctly matched, although it is difficult to check for the vast number of levels without strong emission.

Due to its abundance in the Universe and its strong emission over a wide range of wavelengths, line ratio diagnostics based on \ion{Fe}{xvii} should be very powerful. There are, however, significant differences in the values of these line ratios in the literature, making their use problematic.

The most discussed diagnostic line ratios, following notation from \cite{Parkinson1973}, are the 3C/3D ratios and the 3s/3C ratios, where 3C = $2\mbox{p}^5 \, 3\mbox{d}^1 \,{}^1\mbox{P}_1 \rightarrow$ ground, 15.014\AA ; 3D = $2\mbox{p}^5 \, 3\mbox{d}^1 \, {}^3\mbox{D}_1 \rightarrow$ ground, 15.261\AA; and 3s is the sum of the $2\mbox{p}^5 \, 3\mbox{s}^1 \rightarrow$ ground transitions. The 3C/3D line ratio has long been proposed as a measure of the opacity of the plasma, since the 3C line is sensitive to the opacity of the plasma and the 3D line is less so: comparing the observed ratio with the prediction for an optically thin plasma can be used to obtain the optical depth of the plasma.

Several groups have measured these line ratios using EBIT experiments, including those at NIST \citep{Laming2000} and LLNL \citep{Brown1998, Brown2001, Beiersdorfer2002}. These results have been inconsistent with each other, with calculated values of the ratios for these lines, such as \cite{Chen2008}, \cite{Loch2006} and \cite{Laming2000}, and with astrophysically observed values \citep{Brickhouse2006}. Agreement was frequently no better than a factor of 2 between the different methods.

\begin{figure}
\begin{center}
\includegraphics[width=\linewidth]{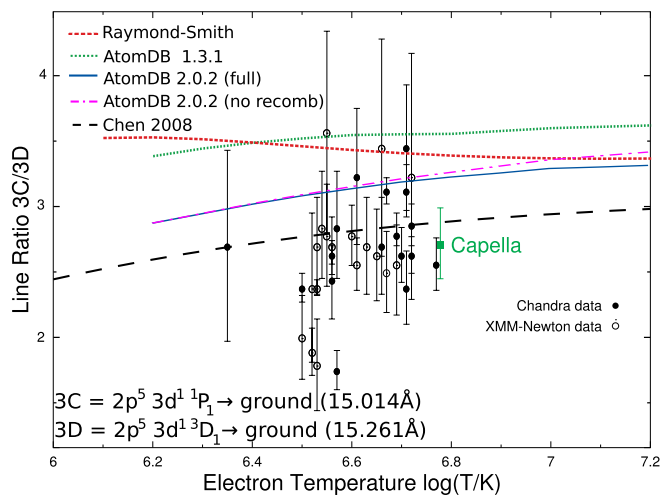}
\caption{\label{fig:fe_xvii} The \ion{Fe}{xvii} 3C/3D line ratio from various calculations, compared with observed data. Solid line: AtomDB v2.0.2; dash-dotted line: AtomDB v2.0.2 excluding state-selective recombination; dotted line: AtomDB v1.3.1; small dashed line: the Raymond Smith model \citep{Raymond1977}; large dashes: \cite{Chen2008}. Various observed values are also taken from a variety of Chandra and XMM-Newton observations (solid and open circles, respectively). Figure adapted from \cite{Chen2008}.}
\end{center}
\end{figure}

Recently the work of \cite{Gillaspy2011} appears to have shown that with careful consideration of the EBIT plasma and instrumental effects such as window absorption, as well as revisiting RR recombination and its populating effect on upper levels, that observed values can agree to within 5 to 20\% of the calculated values. We show in Figure~\ref{fig:fe_xvii}, adapted from \cite{Chen2008}, the 3C/3D line ratio in iron from a variety of astrophysical objects. Also shown are the 3C/3D ratios form the Raymond-Smith code, from AtomDB v1.3.1, from AtomDB v2.0.2, and from \cite{Chen2008}.

In this case, the AtomDB v2.0.2 data is that of \cite{Loch2006}. We note that the values from \cite{Chen2008} are closer to the EBIT measurements and to the mean of the observed values, and therefore that their line ratio values are more likely accurate. However, this calculation is incomplete for the remaining strong transitions of \ion{Fe}{xvii}, and as individual line ratios are a complex result of many different excitation and de-excitation processes, this data cannot be included in AtomDB. We have therefore used the \cite{Loch2006} data in AtomDB v2.0.2, providing a significant improvement over the data from v1.3.1, but this will be revisited when further data becomes available; for now an uncertainty of up to 20\% should be assumed when using values.

As noted by \cite{Doron2002}, the addition of the state-selective recombination rates leads to some suppression of the 3C/3D line ratio at high temperature. They approached this issue by constructing a three ion model to include ionization and recombination, and our results here do not conflict with their findings.

\subsection{Other Ions}
For many other ions, especially those that are not notably strong emitters in the X-ray region, AtomDB v1.3.1 used data from CHIANTI v2.0 \citep{Landi1999}. We have updated the wavelengths, transition probabilities and collision strengths of all data for all ions excluding the H-like \& He-like sequences, and the Ni and Fe L-shell ions to now incorporate the CHIANTI v6.0.1 data \citep{2009A&A...498..915D}, which was the current version at the time of preparation of this data set. The Ni L-shell data remains that of the previous version of AtomDB, which is scaled from the Fe data of Liedahl. We will aim to address this in a future update.

\subsection{Minor Changes}
In addition to these major updates to the database, several minor corrections have been made which have little or no effect on the resulting spectra from AtomDB but which do affect the database. These include the correct use of L and S quantum numbers for all levels where LS coupling is used; consistent formatting of configuration strings, minor corrections to the H-like two-photon transition rate from the $2\mbox{s}^1 \rightarrow 1\mbox{s}^1$ level, including adding the rate for hydrogen and reducing it for all other H-like ions. None of these changes have a noticeable effect on the spectra from AtomDB v2.0.2 compared with AtomDB v1.3.1.

\section{Results}\label{sec:results}

\subsection{Effects of Improved Ionization and Recombination Rates}
\begin{figure}
\centering
\includegraphics[width=\linewidth]{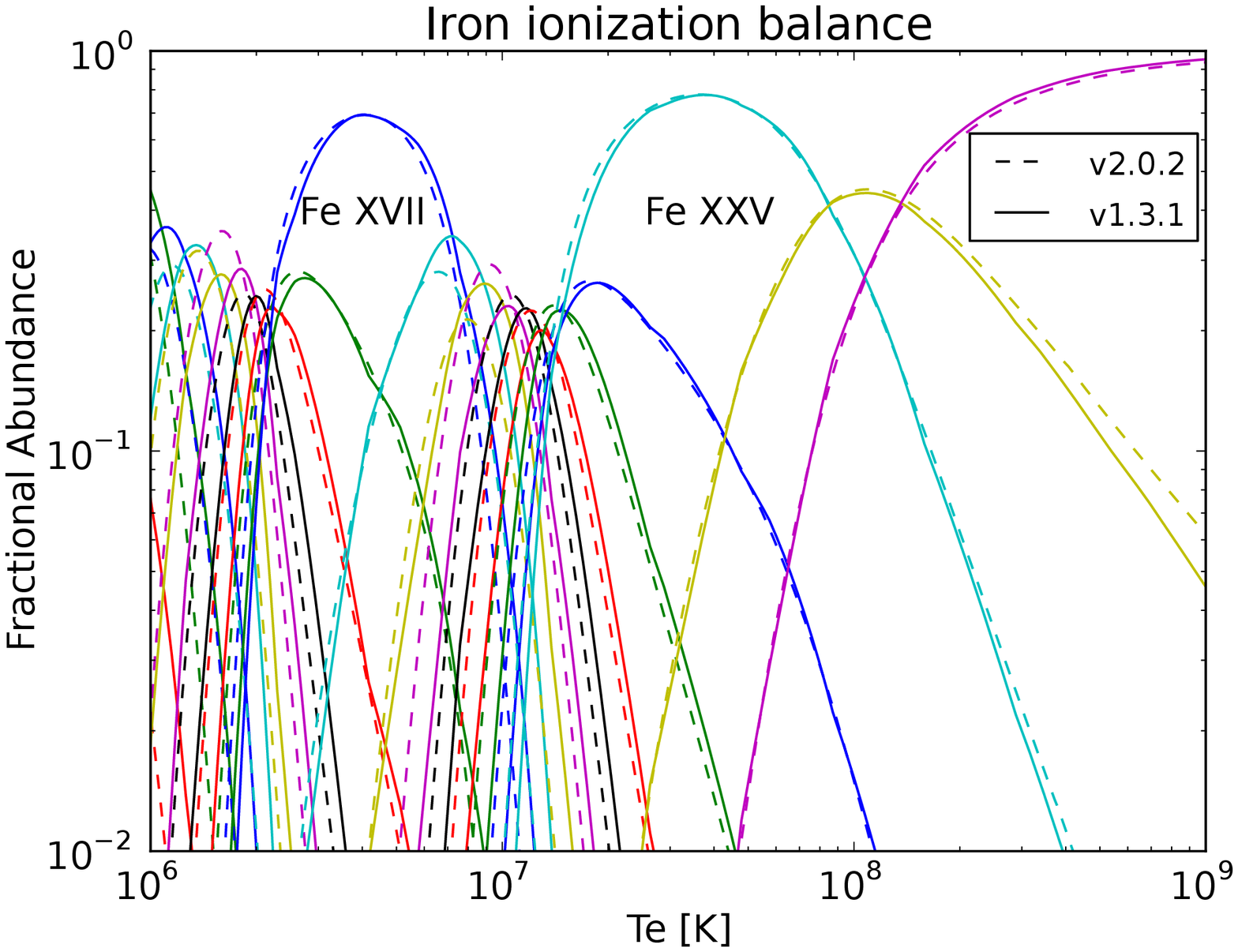}
\includegraphics[width=\linewidth]{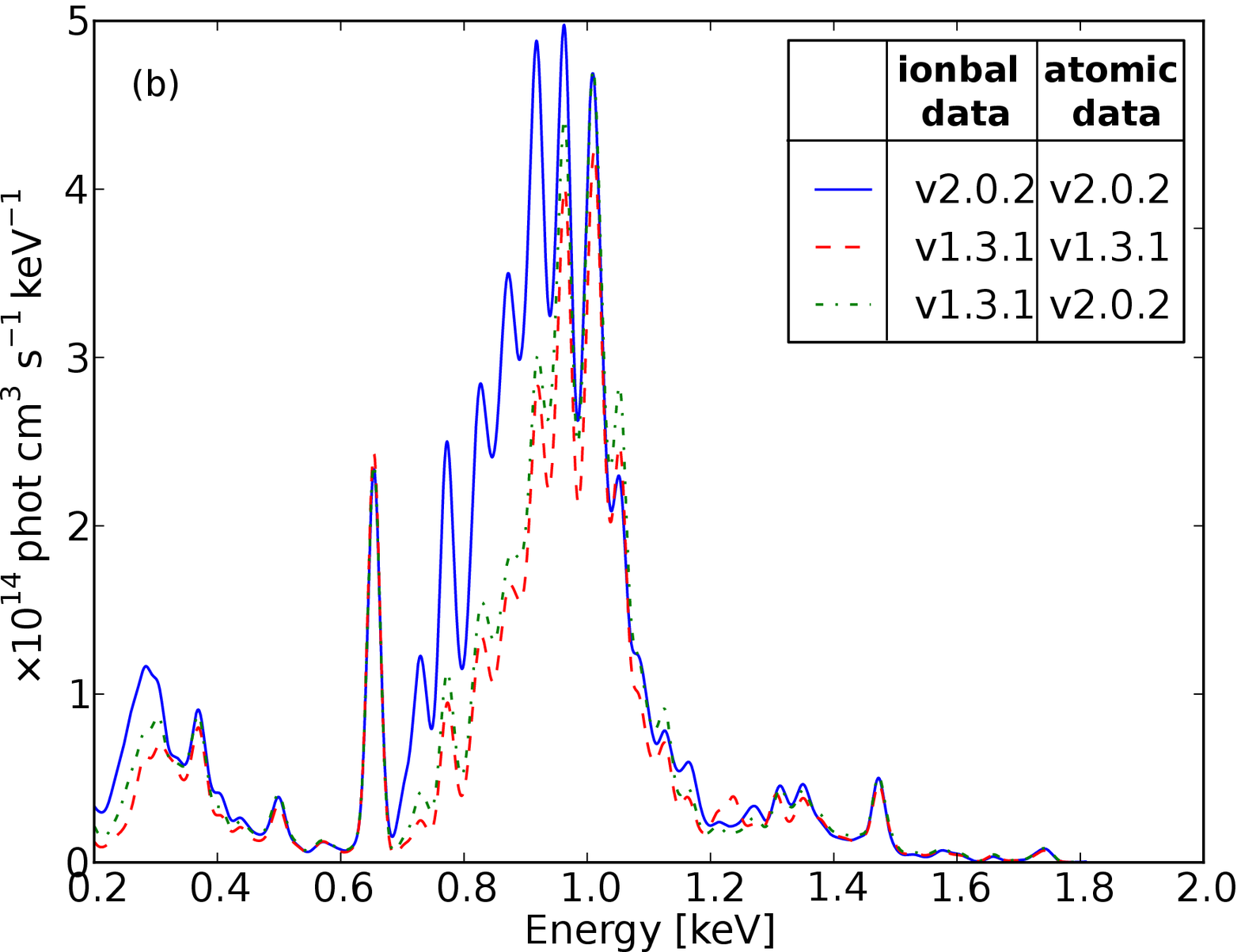}
\caption{\label{fig:ionbal} \textit{Top:} The ionization balance for iron above $10^6$K. The dashed line is the \cite{Mazzotta1998} data used in AtomDB v1.3.1, the solid line is the newer data mostly based on \cite{Bryans2006,Bryans2009} included in v2.0.2. Significant changes are found in the $10^6-10^7$K range. \textit{Bottom:} A comparison of the continuum and line emission from v1.3.1 and v2.0.2 of AtomDB, in a 1keV optically thin plasma with solar abundances \citep{Anders1989}. Lines have been broadened by convolution with Gaussians, with FWHM=0.0235 keV. Also shown is the line emission using all of the atomic data from AtomDB v2.0.2, but the ionization balance from v1.3.1. The ionization balance is the dominant cause of changes in the spectrum, with more Fe$^{+17}$ and Fe$^{+18}$ present in version.}
\end{figure}

Changes in the ionization balance are most significant for heavier elements. Detailed information can be found in the papers of \cite{Bryans2006,Bryans2009}. Figure~\ref{fig:ionbal} shows the change in the ionization balance for iron, including the recombination rates for \ion{Fe}{xix} as specified in Section \ref{sec:newdata:ionrec}. Significant changes can be seen around \ion{Fe}{xvii}, in some cases leading to changes in fractional abundance of up to 30\%.

The most readily apparent spectral changes between AtomDB v1.3.1 and v2.0.2 are due to the change in the ionization balance data. The shift in the ionization balance leads to stronger emission from many of the iron L-shell ions at temperatures between $10^6$ and $10^7$K. In the lower part of Figure~\ref{fig:ionbal}, the line emissivities from AtomDB v1.3.1 are compared to that of AtomDB v2.0.2 for a 1~keV optically-thin, collisionally-ionized plasma, with $N_e=1\mbox{cm}^{-3}$. Also, a combination of the v2.0.2 data with the v1.3.1 ionization balance of \cite{Mazzotta1998} is shown, to highlight the effects of the new ionization balance as opposed to all the other new data. As can be seen, the majority of the change in emission is due to the ionization balance, which leads to increased \ion{Fe}{xviii} and \ion{Fe}{xix} emission in the 0.7 to 1~keV band. The continuum emission in all three cases changes very little, and therefore is not plotted in detail.

\subsection{Cooling Function}
\begin{figure}
\begin{center}
\includegraphics[width=\linewidth]{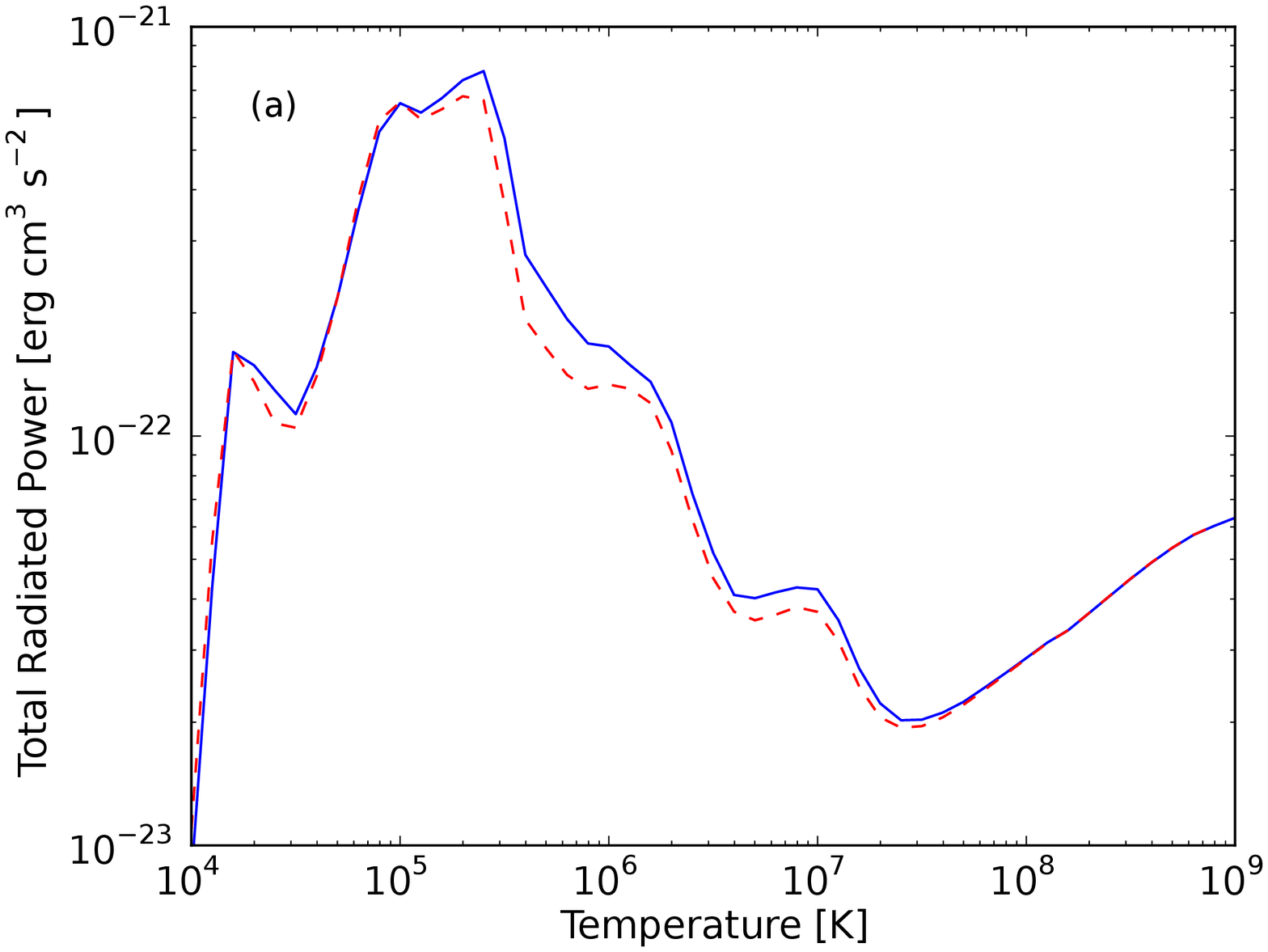}
\includegraphics[width=\linewidth]{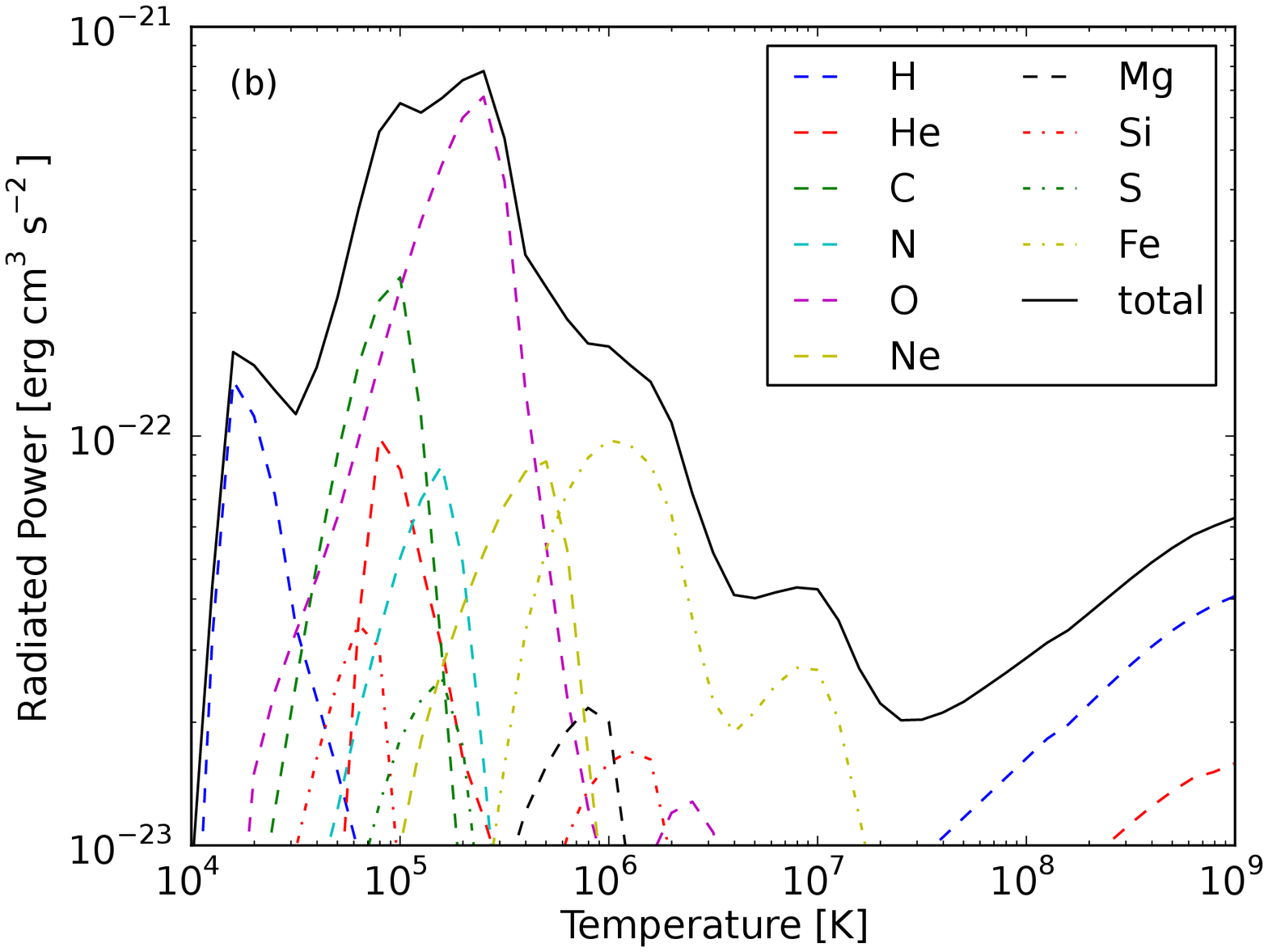}
\caption{\label{fig:cooling} \textit{Top:} the cooling function in the 0.001 to 100keV band from all ions, assuming solar abundances \citep{Anders1989}, comparing the old (v1.3.1, dashed) and new (v2.0.2, solid) AtomDB. \textit{Bottom} The cooling function for AtomDB v2.0.2, broken down by element.}
\end{center}
\end{figure}

Figure~\ref{fig:cooling} shows the total power radiated from all elements given a solar photospheric elemental abundance \citep{Anders1989} and an electron density of 1~cm$^{-3}$. The top figure shows a comparison between the total emission from the old and new versions of the database: as can be seen, there are significant ($> 10\%$) increases in radiation in three different temperature ranges. These are due to changes in the ionization balance for oxygen, neon and iron respectively: in the new data a higher fraction of less-ionized ions is present. These ions radiate more strongly in higher temperature regions, in part due to increased recombination-cascade emission. Figure~\ref{fig:cooling} also shows the radiation separated by element as a function of temperature.

\subsection{Higher $n$-shell Effects}
\begin{figure}
\begin{center}
\includegraphics[width=\linewidth]{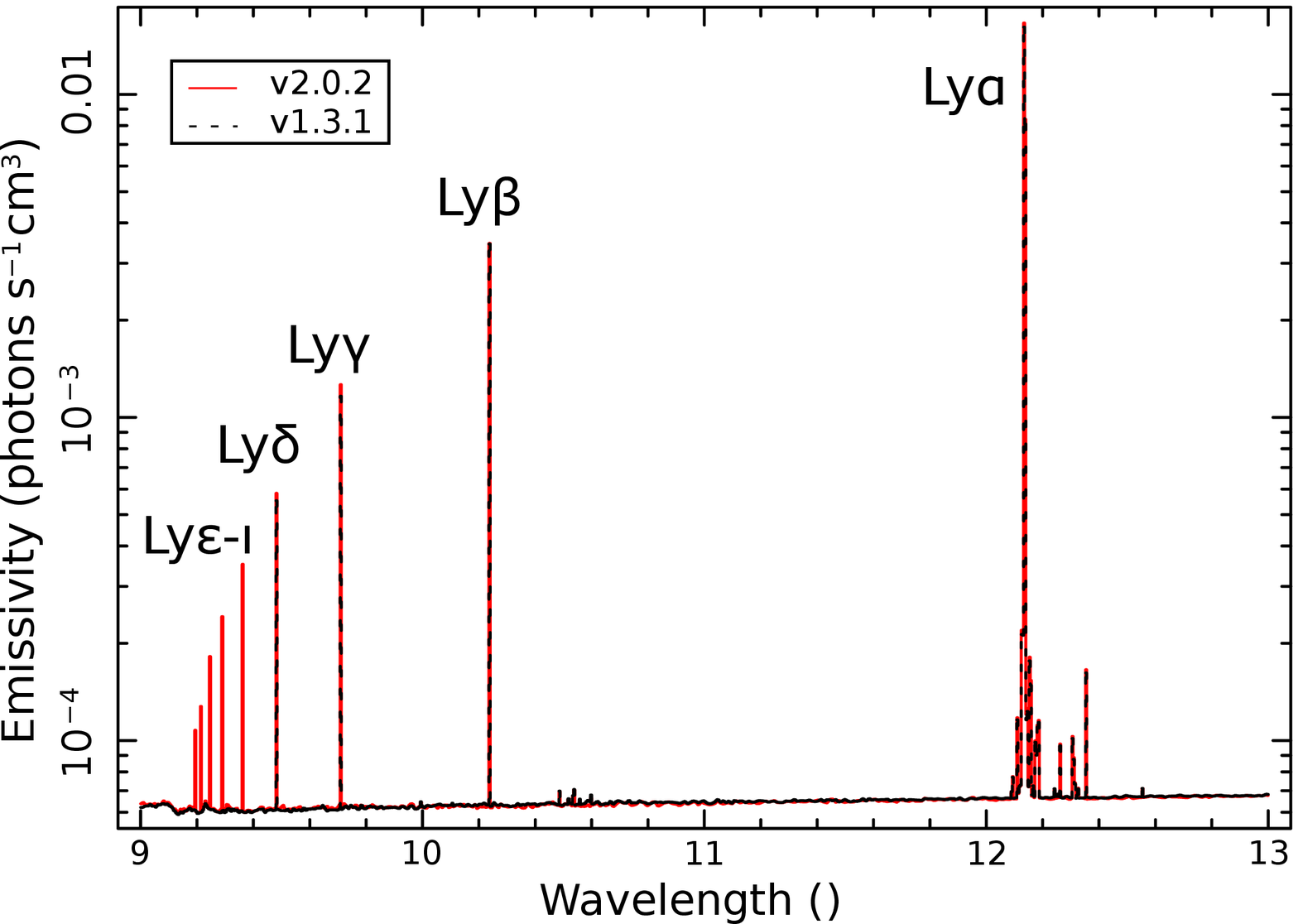}
\includegraphics[width=\linewidth]{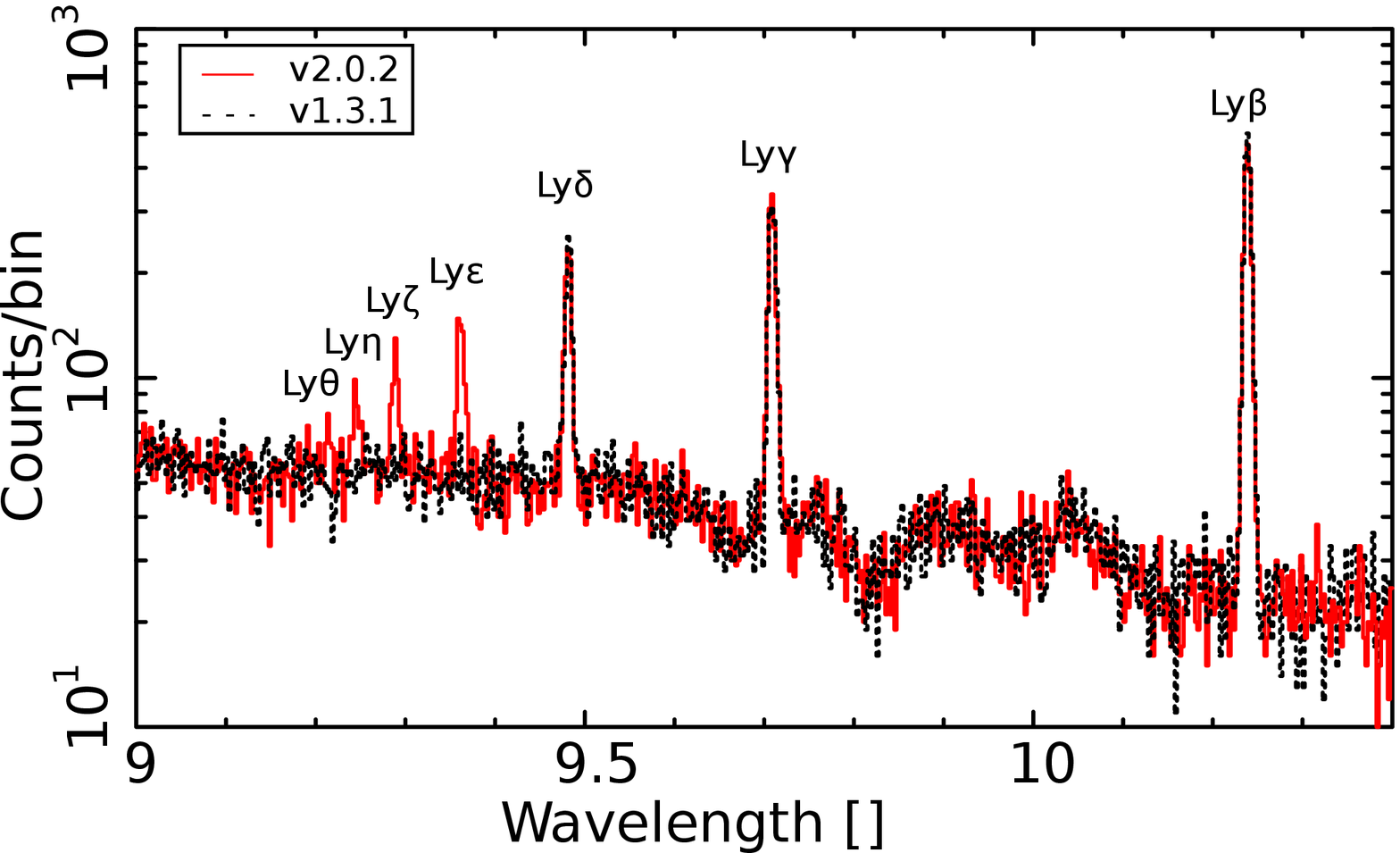}
\caption{\label{fig:h-like_ne} \textit{Top:} the line emissivities for \ion{Ne}{x} at $10^7$K, showing the old data (dashed line) and the new (solid). The lines from the high-$n$ cascade can be seen clearly. \textit{Bottom:} A simulated spectrum of \ion{Ne}{x} from a 100ks Chandra HETG observation, also at $10^7$K.}
\end{center}
\end{figure}

The inclusion of higher $n$-levels for H- and He-like systems has two effects: in addition to new lines formed from these levels, cascade effects to lower $n$ levels can also alter line intensities. Figure~\ref{fig:h-like_ne} shows the emissivities of lines of \ion{Ne}{x} with the higher $n$-shell lines apparent and demonstrates the possibility of either observing these lines, or, more likely, having these lines blend with other lines from Mg XI.

\begin{figure}
\begin{center}
\includegraphics[width=\linewidth]{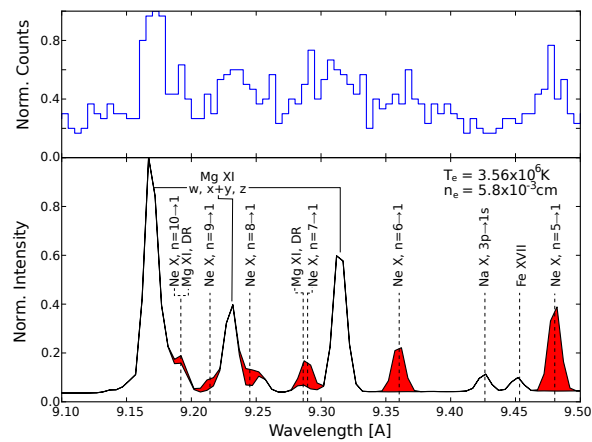}
\caption{\label{fig:h-like_me_mg} \textit{Top:} Summed Chandra HETG spectrum of TW Hydrae \citep{Brickhouse2010} in the Mg XI bandpass. \textit{Bottom:} A simulated AtomDB spectrum with Gaussian broadening applied to each line. Contributions from \ion{Ne}{x} are shaded in red.}
\end{center}
\end{figure}

Figure~\ref{fig:h-like_me_mg}\ shows the \ion{Mg}{xi} bandpass (9-10\AA) of the Chandra HETG spectrum from TW Hydrae \citep{Brickhouse2010}, compared to a simple optically-thin, collisionally ionized plasma spectrum created using AtomDB v2.0.2. Line features are broadened by convolution with Gaussians with FWHM=0.010\AA. Contribution from Ne X to the spectrum is highlighted in red. Solar abundances of \cite{Anders1989} are used throughout, except for Ne (=1.23 solar) and Mg (=0.18 solar). These values are taken from Model A in \cite{Brickhouse2010}, as are the plasma conditions of $T_e = 3.58\times 10^6$K, $n_e = 5.8\times 10^{12}$cm$^{-3}$. The inclusion of the $n > 5$ levels leads to several extra lines being apparent in the spectrum, potentially affecting line ratio estimates depending on the resolution of the instrument in use.

\subsection{Diagnostic Line Ratios}

\subsubsection{He-like ratios}
\label{section:helikeratios}

\begin{figure}
\begin{center}
\includegraphics[width=\linewidth]{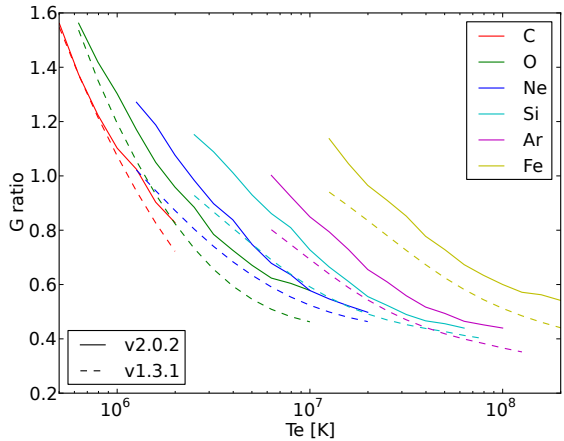}
\caption{\label{fig:gratios} The temperature sensitive G-ratio for selected He-like ions, for both the old (v1.3.1, dashed) and new (v2.0.2, solid) AtomDB.}
\end{center}
\end{figure}

The effect on the G-ratios (Section \ref{sec:helike}) of the new data used in AtomDB v2.0.2 can be seen in Figure~\ref{fig:gratios}. For most ions the G-ratio substantially increases, in some cases leading to a doubling of the implied temperature. At the peak temperatures for each ion, this effect is attributed largely to an increase in the collisional excitation rate to the forbidden and intercombination transition upper levels: this has led to, for example, a 30\% increase in both the forbidden and intercombination line emission for Si XIII. At slightly higher temperatures, above which the $n=2\rightarrow 1$ complex emission has peaked, the recombination and cascade processes have the most significant effect.

Comparisons of the temperature implied by the G-ratio with that implied by the ratio of the Ly-$\alpha/w$, where $w$ is the He-like resonance line are sometimes used to investigate plasma ionization states. The two line ratios measure slightly different temperatures. The G-ratio measures directly the electron temperature of the plasma, due to the fast electron collision rates. The Ly-$\alpha/w$ ratio, however, primarily depends on the ionization balance, and therefore can be used to indicate the ionization temperature of the plasma. If these two diagnostics disagree about the electron temperature, then the plasma is either over- or under-ionized; this is possible since ionization and recombination occur on much slower time-scales than collisional excitation. For a single temperature plasma in ionization equilibrium they will be equal; however, in an object with a broad thermal distribution, the Ly-$\alpha/w$ ratio can be elevated due to Ly-$\alpha$ emission from warmer plasma. 

\begin{figure}
\begin{center}
\includegraphics[width=\linewidth]{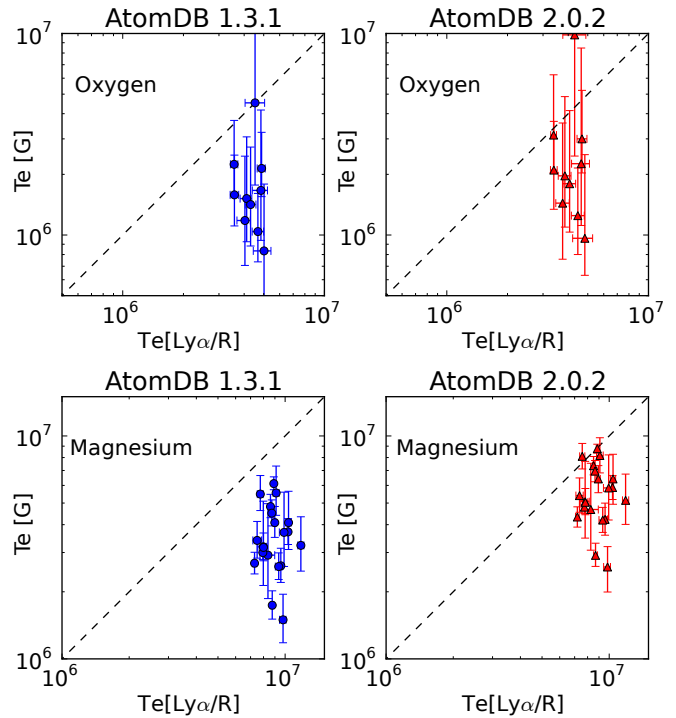}
\caption{\label{fig:testa}A comparison of the temperatures implied by the measured G-ratios and Ly-$\alpha$/resonance lines for Chandra MEG observation data from \cite{Testa2004}, using AtomDB v1.3.1 data (as in the original paper) and the new AtomDB v2.0.2 data. Error bars are based on the error in the line fluxes from \cite{Testa2004}; X-axis errors are very small due to the very strong temperature dependence of the Ly-$\alpha$/w ratio.}
\end{center}
\end{figure}

\cite{Ness2003} encountered this problem when comparing the G-ratio for \ion{Ne}{ix} with the emission measure obtained from iron L-shell ions in Capella: the emission measure peaks at around 6MK, but the G-ratio implied a $T_e$ of around 2MK. \cite{Testa2004} noted that in Chandra observations of a variety of stars, the temperature was consistently underestimated by the G-ratios, sometimes by up to a factor of four. In Figure~\ref{fig:testa} we repeat this comparison, using the data from AtomDB v2.0.2, and the line fluxes from Tables 6 and 7 of their paper. Using the AtomDB v1.3.1 data, we have been unable to reproduce the exact temperatures from their paper, but the same trend is clearly evident.

\cite{Testa2004} and \cite{Ness2003} discuss the causes of some of these discrepancies. The differential emission measure (DEM) of the stars in the survey is generally peaked around $10^7$K (see references in \citealt{Testa2004}). The temperatures of peak emissivity of \ion{O}{vii} and \ion{Mg}{xii} are $\approx 2\times 10^6$K and $\approx 6.4\times 10^6$K respectively. By $10^7$K, the emissivity of the \ion{O}{vii} has dropped two orders of magnitude from its peak. Therefore we would expect the G-ratio from oxygen to be slightly higher than the peak emissivity at $2\times 10^6$K, but not as high as $10^7$K due to the lack of \ion{O}{vii} emission at this temperature. The \ion{Mg}{xi} G-ratio should give a temperature closer to the peak of the DEM at $10^7$K as the emissivity of this ion is still high. Instead, they found that the G-ratio in both cases gave temperatures which were \textit{below} the temperatures of peak emissivity, which suggested a problem in the atomic data.

With the AtomDB version 2.0 data, we find that the G-ratio now indicates temperatures which lie on average at or above the peak emissivity temperature for each ion (some spread remains due to the different DEMs of different objects). In addition, the discrepancy between the two temperature measurement methods has been reduced from 60\% to around 35\%.

\begin{figure}
\begin{center}
\includegraphics[width=\linewidth]{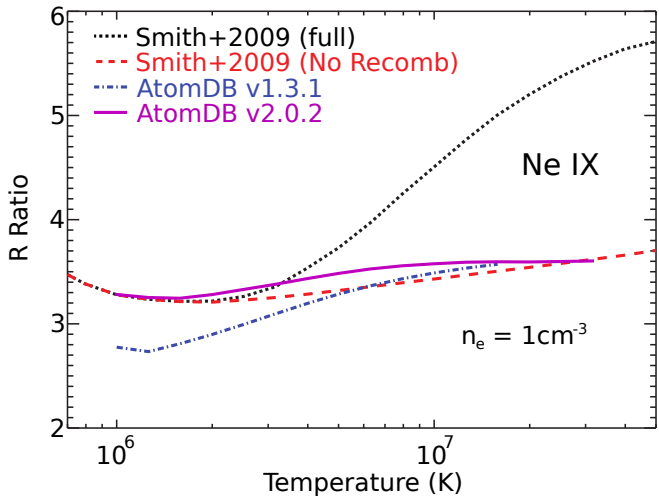}
\caption{\label{fig:neix_fixed}The density diagnostic R-ratio from AtomDB v1.3.1 (dash-dot), the AtomDB v2.0.2 data (solid), and the data from \cite{Smith2009} both with (dotted) and without (dashed) recombination.}
\end{center}
\end{figure}

\cite{Smith2009} also investigated the G- and R-ratios for the case of \ion{Ne}{ix}. They used AtomDB v1.3.1 data, however they updated it to include the more recent collisional data of \cite{Chen2006} (discussed in section ~\ref{sec:helike}) and the recombination data from the \cite{Badnell2006} collection - the same data included in the release of AtomDB v2.0.2. The authors found a significant variation with temperature in the R-ratio, which is the ratio of the forbidden to the inter-combination lines of He-like ions. This temperature dependence is unusual: the G-ratio is typically a widely used density diagnostic with little temperature variation. We do not find any temperature dependence here, despite using almost identical data (the small differences in the collisional excitation calculations have almost no effect). This was traced to an error in the \cite{Smith2009} paper when connecting the energy levels in the recombination data to the excitation data. Figure~\ref{fig:neix_fixed} shows the corrected R-ratio as a function of temperature.

\section{Summary and Future Plans}
We have presented the latest version of AtomDB, the first major update since 2001. Nearly every single piece of data in the database has been updated, with many new ions added. This is the result of a comprehensive evaluation of the previous data, and assessment of its replacement, addressing many of the known issues in the previous version. New recombination data significantly alter the ionization balance, and therefore the emissivities, of many lines; new collisional data for H- and He-like ions make significant improvements to common temperature diagnostics.

The new data are now available on-line at www.atomdb.org, and are also available through spectral fitting packages such as XSPEC \citep{Arnaud1996}, ISIS \citep{Houck2000} and Sherpa \citep{Freeman2001}.\footnote{After the release of v2.0.0, errors were discovered affecting the radiative recombination continuum and the autoionization rates. This led to the release of v2.0.1 and v2.0.2 of AtomDB. All data in this paper refer to the corrected v2.0.2 data.}

There are several improvements already planned for the next release of AtomDB. We will include final state resolved DR rates for the remaining ions in the database for which the data exist. We will largely target non-equilibrium ionization plasmas, with new inner-shell excitation data, as well as fluorescence line data. Work is also ongoing to document and release the APEC collisional ionization code, which will allow users to generate the non-equilibrium higher density plasma models.

We wish to thank Paola Testa for discussion involving the line diagnostic issues. The authors gratefully acknowledge funding from NASA ADP grant \#NNX09AC71G.

\bibliography{atomdb200.bib}{}
\bibliographystyle{apj}

\end{document}